\documentclass{amia}
\usepackage{graphicx}
\usepackage[labelfont=bf]{caption}
\usepackage[superscript,nomove]{cite}
\usepackage{color}

\usepackage{todonotes}

\usepackage{cleveref}
\usepackage{hyperref}
\usepackage{url}
\usepackage{soul}

\usepackage{amsmath,amssymb,amsfonts,mathrsfs}
\usepackage{booktabs}
\usepackage{subcaption}

\begin{document}

\title{Embedding Electronic Health Records for Clinical Information Retrieval}

\author{Xing Wei, MSc$^{1}$, Carsten Eickhoff, PhD$^{2}$}

\institutes{
    $^1$ETH Zurich, Zurich, Switzerland; 
    $^2$Brown University, Providence, RI\\
}

\maketitle

\noindent{\bf Abstract}

\textit{Neural network representation learning frameworks have recently shown to be highly effective at a wide range of tasks ranging from radiography interpretation via data-driven diagnostics to clinical decision support. This often superior performance comes at the price of dramatically increased training data requirements that cannot be satisfied in every given institution or scenario. As a means of countering such data sparsity effects, distant supervision alleviates the need for scarce in-domain data by relying on a related, resource-rich, task for training.} 

\textit{This study presents an end-to-end neural clinical decision support system that recommends relevant literature for individual patients (few available resources) via distant supervision on the well-known MIMIC-III collection (abundant resource). Our experiments show significant improvements in retrieval effectiveness over traditional statistical as well as purely locally supervised retrieval models.}

\section*{Introduction}
Electronic Health Record (\emph{EHR}) data is becoming more and more accessible due to the digitization of modern hospitals. Databases such as MIMIC-III~\cite{MIMIC-III} provide detailed information related to a patient, and such datasets give a unique opportunity for improving biomedical literature retrieval qualities~\cite{greuter2016eth} as well as downstream tasks such as question answering~\cite{galko2018biomedical} or forecasting adverse events~\cite{grnarova2016neural,meyer2018machine} . 

Meanwhile, making use of EHR data, especially biomedical notes in raw natural language, can be quite a challenging task~\cite{kuhn2016implicit}, as the semantics of biomedical text are difficult to represent in machine understandable forms. There are some information retrieval tasks that simulate the scenario of medical literature retrieval. In the TREC Clinical Decision Support task~\cite{roberts2015overview}, for example, a medical note including summary and description of a patient is provided, and the goal is to retrieve relevant documents. Such tasks differ from the traditional information retrieval tasks in that the query is much longer, and typically contains multiple paragraphs. 

Although word embeddings are already proven to be effective in representing a word's semantic meanings in dense vector, they are not sufficient when the text becomes longer. And just taking the mean or the maximum of word embeddings in sentences/paragraph doesn't give good results. 

In this paper, we aim at finding a good embedding of a piece of text. To do this, in the first phase we utilize the diagnostic ICD codes and medical notes from the MIMIC-III database, using pre-trained word embeddings, we train a convolutional neural network to predict patients' diagnostic ICD codes from the raw medical note text. The hidden layer of the fully connected network is then extracted and used as the dense representation of the text. The reason behind this method is that word embeddings can give semantic meanings of terms in text, while the context of a paragraph can be captured by the convolution and max pooling layers in the convolutional network. 

In the second phase, we apply our embedding model to biomedical article retrieval tasks. To achieve this, we first apply our model to get a vector representation of each of the paragraphs in query and candidate documents, we then feed the obtained vectors into a recently-proposed deep relevance matching model (\emph{DRMM}), and this model will output a ranking score for each query-document pair. We use cosine similarity between query and candidate document embeddings as our ranking score. 

Our trained embedding model can embed any sentence or paragraph of text into a feature vector. We do not train the embeddings directly from retrieval tasks because of the lack of sufficient training data for our model. Instead, we add a classification step to train embedding model because such kind of data is abundant. The classification problem serves as a proxy to training effective embeddings due to the lack of retrieval data. 

In our experiment, the embedding model gives better feature vectors for ICD code prediction tasks, which shows the fact that our model act effectively as a good dense representation capturing import aspects of the underlying text. And our embedding applied to DRMM gives a better performance over baseline retrieval algorithms. We also try our embedding framework on more general information retrieval tasks such as the Tipster dataset, although the performance of our model is not as well as the term-based DRMM model, it still gives a better result than the traditional IR algorithms such as BM25. In our experiments, the best result is attained by an ensemble method that combines the term-level DRMM, the paragraph-level DRMM and cosine similarity of document embeddings. 

\section{Related Work}
In this section, we give a quick overview of related work in the field of text embedding and information retrieval, comment on their advantages and shortcomings and compare to our method. 

% Head 2
\subsection{Vectorization for text or EHR}
\label{sec:related-works-text-vec}

There exists a lot of work on the vectorization of a piece of text. Classical methods include bag-of-words, tf-idf, etc. The problem with such approaches is that the vector is very sparse, and does not contain much information on the semantics level. On the other hand, word embeddings such as Word2Vec \cite{mikolov2013efficient} and GloVe \cite{pennington2014glove} have proven to be effective in representing the semantics at word level with a distributed representation. But when it comes to representing the semantics at sentence or paragraph level, simply averaging or taking the maximum of word vectors is not sufficiently effective any more.  

% Head 3
\subsubsection{Paragraph vectors}

\begin{figure}[!htb]
  \includegraphics[width=4in]{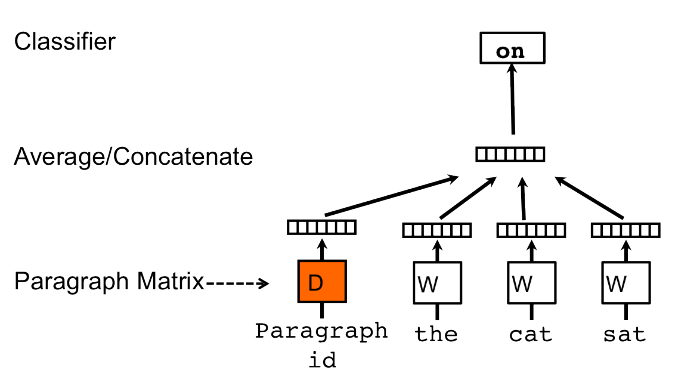}
  \centering{}
  \caption{Framework of learning paragraph vector. \cite{DBLP:journals/corr/LeM14}. The paragraph vector along with other word vectors in context are used to predict the next word.}
  \label{fig:paravec}
\end{figure}

Le et al.\ \cite{DBLP:journals/corr/LeM14} propose an unsupervised paragraph embedding method, this involves concatenating the (to be trained) paragraph vector and (pre-trained) word vectors in a sliding window, and predicting the next word. The paragraph vector is trained using a gradient descent algorithm. The framework is shown in Figure \ref{fig:paravec}. 

This algorithm incorporates the power of word vectors in representing semantics and takes the word order and context into account. According to the paper, paragraph vectors can capture the current context and serve as a memory of the topic of the paragraph. The problem with this approach is that, when given a piece of text, the paragraph vector has to be trained using gradient descent. As the convergence of the gradient descent might take time, this approach might not be suitable for handling a large amount of texts. Also when the paragraph becomes very long, the paragraph vector will be averaged as the sliding window moves and would become less accurate. 

\subsection{Deep patient vectors}\label{ssec:deep-patient}

When treating electronic health records (EHR), medical concepts can be extracted using unsupervised tools such as QuickUMLS by Soldaini and Goharian \cite{quickUMLS}. Given a medical note, running QuickUMLS on it gives us a list of medical concepts extracted from this note. 

\begin{figure}[!htb]
  \includegraphics[width=5in]{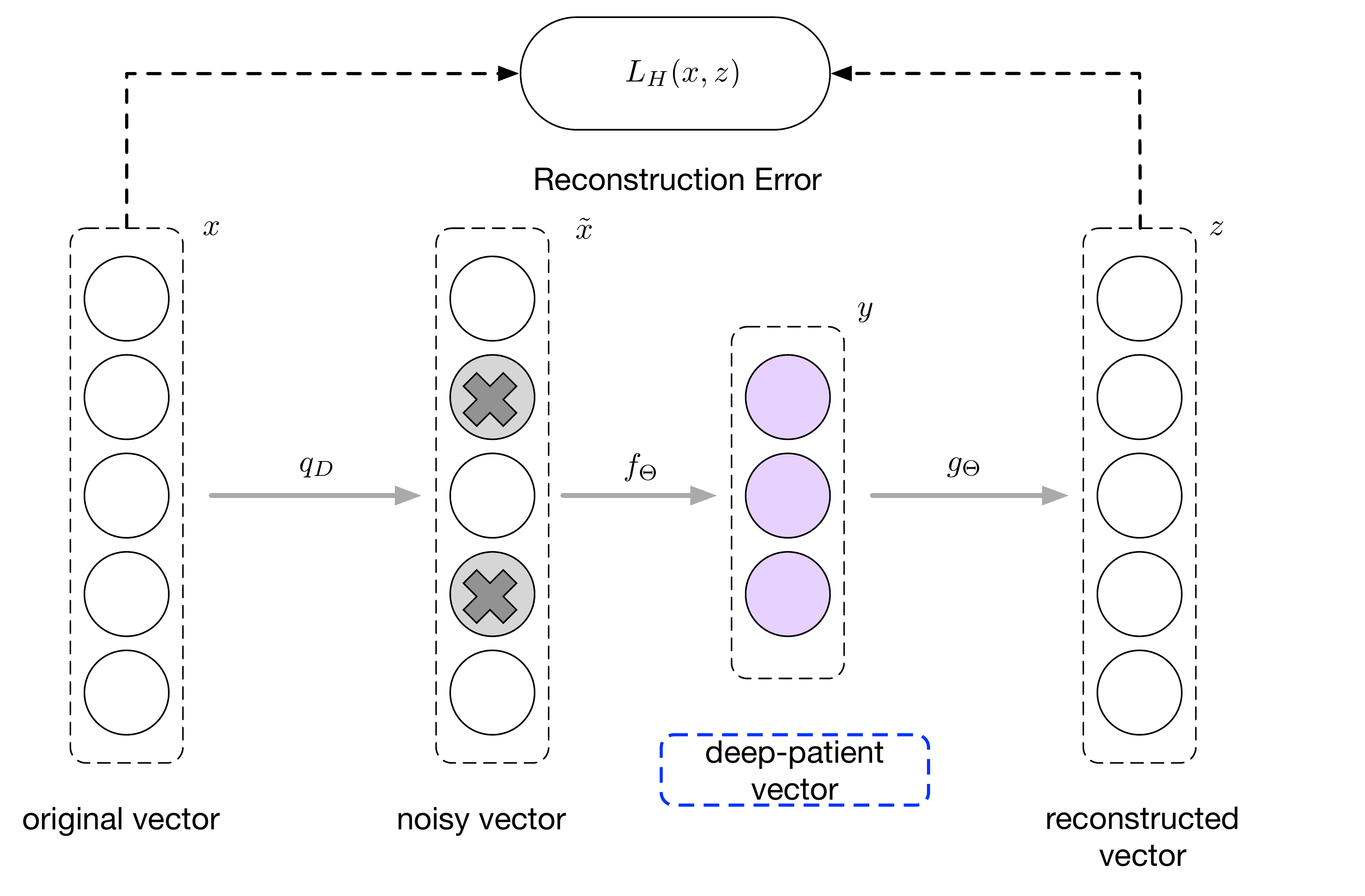}
  \centering
  \caption{Training of a denoising autoencoder}
  \label{fig:denoiseAE}
\end{figure}

The raw result of QuickUMLS might still be very sparse as there are a lot of possible medical concepts. To solve this, Miotto et al.\ \cite{deep-patient} proposed an unsupervised algorithm to extract a denser vector representation. They added some noise to the raw (sparse) binary medical concepts feature vector, and trained an autoencoder to recover the feature vector without noise. This denoising autoencoder is shown in Figrue \ref{fig:denoiseAE}. 

\begin{figure}[!htb]
  \includegraphics[width=5in]{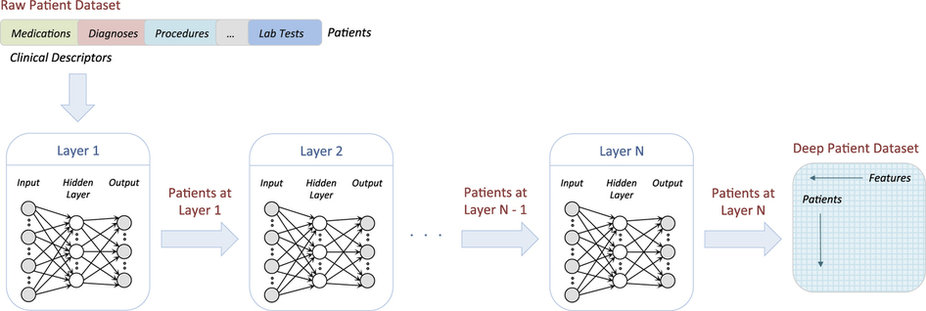}
  \centering
  \caption{Diagram of the unsupervised deep-patient \cite{deep-patient} feature learning pipeline. The input is raw binary feature vector of medical concepts from QuickUMLS. Parameters of autoencoders are the same for all layers. The hidden layer of last autoencoder is extracted as deep patient embedding vector.}
  \label{fig:deep-patient}
\end{figure}

After training such an autoencoder, they stack several layers of the same autoencoder, and the hidden layer of the last autoencoder is used as a "deep-patient" feature vector. This framework is shown in Figure \ref{fig:deep-patient}. 

According to their paper, the deep patient embedding outperforms traditional vectorization algorithms on tasks such as disease classification and patient disease tagging. The raw input to this model can be seen as a bag-of-concepts representation of the notes, applying autoencoders to the input generates dense, low-dimensional ($dimension=500$) feature vectors. 

While this representation might be generally useful to represent a \emph{patient}, this might not be appropriate for biomedical literature retrieval, because the QuickUMLS might not work well on the content of academic articles. And this method is limited only medical notes, because a list of extracted medical concepts has to be extracted as input. Our goal in this article is to create a more general model that can embed directly a piece of text into a dense vector. 

\subsection{Information retrieval}
\label{sec:related-works-IR}

Traditional information retrieval algorithms usually apply term weighting and scoring, and use term exact matches as important indicator of relevance. Algorithms such as BM25 \cite{Manning:2008:IIR:1394399} or tf-idf can work fairly well for queries of relatively short length. When query becomes a fairly long piece of text, the semantics of text become important, and semantic similarity might be more indicative than simple term occurrence matching. To capture semantic similarity, embedding techniques should be involved. 

\subsubsection{Deep Relevance Matching Model}
\label{ssec:drmm}

\begin{figure}[!htb]
  \includegraphics[width=5in]{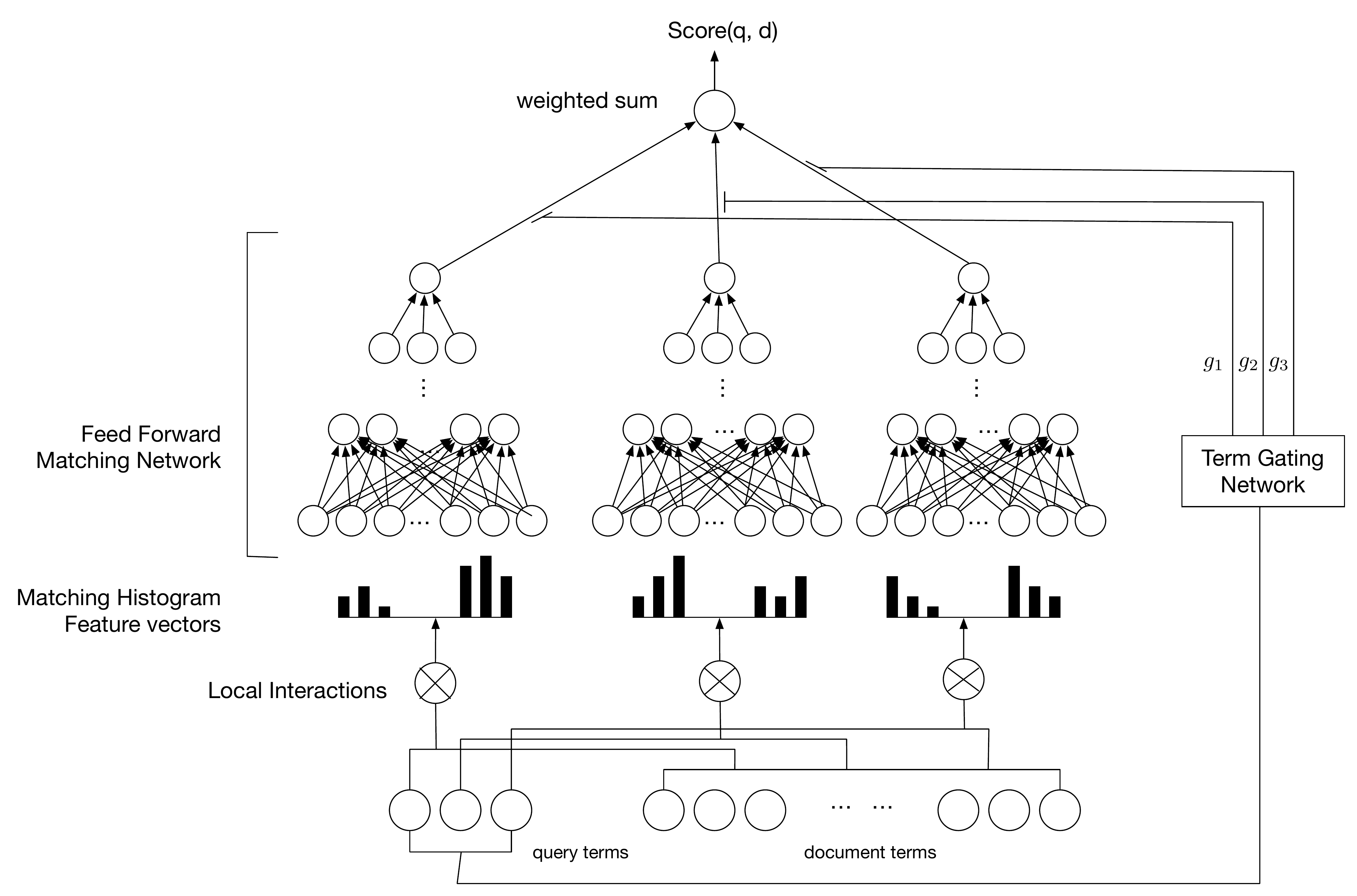}
  \centering
  \caption{Architecture of Deep Relevance Matching Model}
  \label{fig:drmm}
\end{figure}

Guo et al.\ \cite{Guo:2016:DRM:2983323.2983769} proposed in a novel deep relevance matching model for ad-hoc retrieval. To give a score to a query/document pair, the use the relevance matching at the query term level. More specifically, in their model they utilize pre-trained word embeddings. For each query term, they calculate the cosine similarity of this query term with each term in candidate document. A histogram of the cosine similarities is then fed to a feed forward network to generate a score of this query term and candidate document. They generate a score for each term in the query, and weight the scores using the softmax of each query term's inverse document frequency (scaled by a facter $w$). This structure is depicted in Figure \ref{fig:drmm}. 

To train this model, they used a hinge loss of the difference of scores between a relevant document and a non-relevant document, the positive-negative pairs are generated from training qrels. As the whole network is not very big (just a few hundreds of parameters), training of the DRMM model can be quite fast. 

Their experiment on Robust-04 and ClueWeb-09-Cat-B dataset showed that DRMM gives better retrieval results than the BM25 baseline. In our experiment, this model works also well on 2016 TREC-CDS datasets. Although the model involved semantics of words, this model is restricted only to term-level semantics, and we expect a paragraph-level embedding would make the retrieval result even better. 

\subsection{Summary}

The embedding models introduced above are mainly unsupervised, this has the advantage of not requiring labeled data which is often expensive to get. The problem with paragraph vector method is that it does not scale well when treating long paragraphs or treating a large amount of texts. The deep patient embedding is only limited to medical notes, moreover, a QuickUMLS run has to performed to get sparse raw input to deep patient model. 

Our embedding model, in contrast, is supervised: we need tagged texts as training data, and our embedding model is trained to predict labels from texts. In fact, datasets of labeled texts is quite abundant, thus enabling us to train the embedding model with a large amount of parameters. And we do not train the embedding directly from retrieval data because the available training qrels is not as abundant as labeled texts. 

The traditional retrieval algorithms such as BM25 focus on exact matching of query and document terms, and involve term weighting based on term frequency or document frequency. Although BM25 is a highly effective model for typical retrieval tasks where query typically contains just a few terms, it becomes less effective in ad-hoc retrieval tasks, when query becomes much longer. Moreover, semantically close but different terms can not be considered as matching each other in BM25. 

The DRMM model overcomes the semantic matching difficulty by leveraging pre-trained word embeddings. The histogram of query-doc term similarities is a better feature as semantic matching than just exact matchings, the scoring logic is handled by a feed forward neural network, a model than can generate complex scoring functions. And the parameters are learned on training relative/non-relative pairs generated from qrel. Experimental results show that DRMM outperforms traditional retrieval models as well as other state-of-the-art deep matching models. 

The limitation of DRMM lies in the fact that the input is still a bag-of-words representation, word ordering is totally discarded, and contextual semantics can not be utilized by the model. In this article we try to fit our embeddings into the DRMM model, this way the semantics at paragraph level is taken into account, and we expect this contextual similarity to boost the retrieval performance.

\section{Method}\label{chapter:methods}

In this section, we describe our methods of text embedding and how to plug it in the DRMM model for information retrieval. And we provide some implementation details of the models. 
\subsection{Convolutional Neural Networks for ICD code prediction}
\label{sec:methods-cnn}

To train a model for ICD code prediction from raw text notes, we make use of pre-trained word vectors so as to take semantics of words into consideration. In our model, we test with both GloVe vectors trained on Wikipedia texts and Word2Vec vectors trained on biomedical texts. The dimension of embedded word vectors are both $D=200$. 

Starting with the raw text of medical notes, after tokenization, we put the embedding word vector of each token together to get an embedding matrix of shape $D \times n$. The text tokens sequences are padded or truncated to a fixed length of $n$. Then the embedding matrix $X$ is a concatenation word vectors $x_i$: 

$$X = [x_1, x_2, ..., x_n] $$

A convolution of $k$ filters of width $d$ words is applied to the input embedding matrix, producing $k$ feature maps, each of length $n-d+1$. The activation function of convolution layers are rectified linear (ReLU) units. More formally, let $W \in \mathbb{R}^{d \times D}$ be the weight matrix of one kernel, and let $y$ be the resulting feature map vector. The $i$th component of $y$ is computed as shown in Equation \ref{eq:cnn-featuremap}. 

\begin{equation}
\label{eq:cnn-featuremap}
y_i = ReLU(W \cdot X_{i:i+d}+b)
\end{equation}

After each convolution layer, a max-pooling layer of width $d'$ is applied on feature maps, the number of filter maps is thus reduced, and only strong feature map activations are preserved. Let $z$ be the output of the max-pooling layer, the $k$th component of $z$ is calculated as in Equation \ref{eq:cnn-maxpooling}. 

\begin{equation}
\label{eq:cnn-maxpooling}
z_i = max(y_{i \times d'}, ..., y_{i \times d' + d'-1})
\end{equation}

We apply several such convolution-maxpooling layers. The last feature map is flatten and passed to a fully connected layer with sigmoid as activation function. The output of our network is a vector of probability $p \in [0,1]^K$ ($K$ is the number of labels to predict), $p_i$ represents the probability that the input instance has label $i$. The ground truth labels are encoded as a binary k-hot vector $q \in \{0,1\}^K$. 

We do not use softmax as activation function for the fully connected layer, because in our case the target has more than one label, we choose sigmoid as final activation function. The loss function is pairwise binary cross entropy as shown in Equation \ref{eq:cross-entropy}. 

\begin{equation}
\label{eq:cross-entropy}
Loss(p,q) = \overset{K}{\underset{i=1}{\sum}}q_i\log p_i
\end{equation}

\begin{figure}[!htb]
  \includegraphics[width=5in]{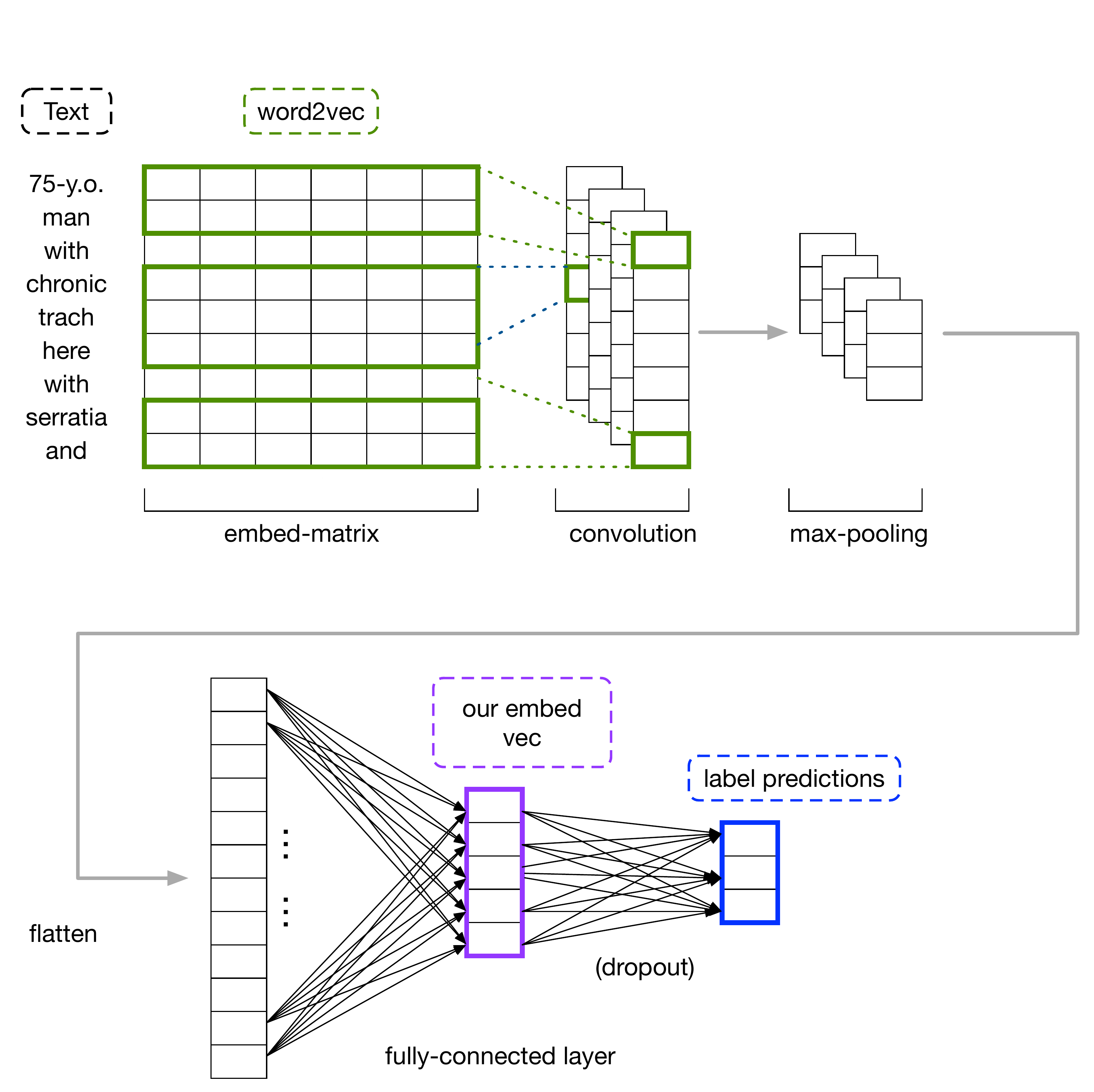}
  \centering
  \caption{Convolutional neural network for text label prediction and text embedding. }
  \label{fig:cnn-embedding}
\end{figure}

To regularize our model, a dropout layer (with parameter $p=0.5$) is applied before the final layer. Once the model is trained, we can feed any text, and take the penultimate layer as the embedding vector. Equation \ref{fig:cnn-embedding} demonstrates how the model works.

\subsection{Embedding Vectors applied to retrieval models} 
In a general setting, given a query $q = \{w_i^{(q)}, i=1,...,M\}$ and a document $d$, the score is a weighed sum of relevances between query word $w_i^{(q)}$ and $d$: 

\begin{equation}
Score(q,d) = \overset{M}{\underset{i=1}{\sum}}W_i \times Rel(w_i^{(q)}, d)
\end{equation}

\subsubsection{DRMM at term level}
As described in Equation \ref{ssec:drmm}, the DRMM is a semantic matching model at term level. The semantic relevance between a query term and a candidate document is characterized by a histogram of the cosine similarities between the word embedding of this query term $w_i^{(q)}$ and all terms in document $d$. And each score of query term and candidate document is weighted by a gating function that takes the IDF of query term into consideration. 

More formally, we denote $\otimes$ as the interaction between query term and all document terms, resulting in a sequence of cosine similarities: 

\begin{equation}
w_i^{(q)} \otimes d = \{ cosine\_sim(w_i^{(q)}, w_j^{(q)})| w_j^{(q)} \in d\}
\end{equation}

And we denote $h()$ as the function that turns a sequence of cosine similarities into a histogram of $B=30$ bins. Then the histogram feature vector $z_i$ can be written as in Equation \ref{eq:hist-vec}. 

\begin{equation}
\label{eq:hist-vec}
z_i = h(w_i^{(q)} \otimes d)
\end{equation}

In the implementation, one bin in the histogram is used for counting exact matches. A logarithm is applied on histogram counts to reduce the range and to allow the model to more easily learn multiplicative relationships. 

This histogram feature vector $z_i$ is then fed to a feed forward network of $L$ layers. The final layer output is the term-document relevance score:

$$Rel(w_i^{(q)}, d) = z_i^{(L)}$$ 

In the other part, a gating network generates for each query term $w_i^{(q)}$ a weight $g_i$, in this case, the weight is the softmax function of the scaled (by a parameter $w$) IDF of query terms: 

\begin{equation} 
g_i = \frac{\exp(w x_i)}{\overset{M}{\underset{j=1}{\sum}}\exp(w x_j)}
\end{equation}

The final score for query and document is a weighted sum of $z_i^{(L)}$: 

\begin{equation}
Score(q, d) = \sum{g_i z_i^{(L)}}
\end{equation}

We use qrels to train the DRMM, for each query and generate a list of relevant documents and non-relevant documents from the qrel. Training data is generated as triplets in the form \texttt{query\_id, rel\_docid, nonrel\_docid}. 

Our goal is to have a model that scores high for a relevant document (denoted $d^+$) and scores low for a non-relevant document (denoted $d^-$). The loss is a hinge loss with margin $\Delta =0.1$ between the scores: 

\begin{equation}
Loss(q, d^+, d^-) = max(0, \Delta-Score(q,d^+)+Score(q, d^-))
\end{equation}

One implementation detail in the training is that, to avoid queries that have many relevant documents dominate the training, we over-sample training triplets for queries with fewer relevant documents and sub-sample triplets for those with more relevant documents.  

\subsubsection{DRMM at paragraph level: plugging cnn-embedding vectors}
In the original DRMM model, the document is in a bag-of-word representation, although term semantics are considered via word embeddings, the semantics of entire phrases or sentences are not taken into account. 

Because of the convolution-maxpooling structure of our embedding model described in Section \ref{sec:methods-cnn}, it takes naturally the context in text into account. We can thus split the query and candidate documents into paragraphs, and replace the word embedding with our convolution embeddings of paragraphs. 

More formally, queries and documents are split into paragraphs $p_i$: $q = {p_i^{(q)}, i=1,...,M'}$. For a paragraph $p_i$, we feed it to our CNN embedding model to turn it into a vector $x_i=CNN\_embed(p_i)$. The interaction between a query paragraph $p_i^{(q)}$ is at paragraph level: 

\begin{equation}
p_i^{(q)} \otimes d = \{ cosine\_sim(p_i^{(q)}, p_j^{(q)})| p_j^{(q)} \in d\}
\end{equation}

The histogram feature vector is then generated in the same manner as before: 

\begin{equation}
z_i = h(p_i^{(q)} \otimes d)
\end{equation}

One thing to notice on the histogram feature vector is that this time no bin will be used for counting exact matches -- there will not be any exact matching of paragraphs this time. The term-level relevance score now becomes at paragraph level: 

$$Rel(p_i^{(q)}, d) = z_i^{(L)}$$ 

As for the term weighting, we have two options: uniform weighting, or choose the maximum idf among all words in a query paragraph as an "idf" score for the paragraph. This weighting score is meant to give higher importance for paragraphs that are contains richer information, but in practice, the uniform weight seems to work better. A possible reason might be that as queries usually contain just a few paragraphs, and when the number of paragraphs is small, it is hard to say which paragraph is more important than another.

Finally, we can plug our embedding model into the DRMM framework, and the model is now a relevance/semantic matching at paragraph level. 

\subsubsection{Direct semantic similarity scoring}
As the embedding vector itself already contains compact representation of the text semantics, one more direct way to retrieve documents is by taking just the cosine similarity between embedding of query and candidate document. 

Let $v_q$ and $v_d$ be the output when feeding query / candidate document to our embedding model, the score is simply a cosine similarity between $v_q$ and $v_d$: 

\begin{equation}
\begin{split}
v_q &= CNN\_embed(q) \\
v_d &= CNN\_embed(d) \\
Score(q,d) &= cosine\_sim(v_q, v_d)
\end{split}
\end{equation}

The advantage with this method is that it doesn't need any training, but requires a really good embedding that can contain sufficient semantic information on text. 

\subsection{Summary}
In this section, we have described our main method both for embedding and for retrieval tasks. The embedding model is trained for the purpose of label prediction. We can not train it directly for retrieval because of the lack in training qrels. The effectiveness of our embedding is described in the next section. 

Our embedding plugged into DRMM solves the weakness of bag-of-words representations of text. We expect this change to improve the retrieval performance in the scenarios where query is a whole contextual piece of text rather than just a few words. 

\section{Data}
In this section, we give some introduction to the datasets that we use to run experiments on. More concretely, we use MIMIC-III (\cite{MIMIC-III}) database for our embedding model training, and TREC Clinical Decision Support Track dataset for information retrieval. We also test our methodology on more general datasets other than biomedical information retrieval: we use RCV1 dataset for embedding model training, and Tipster dataset for retrieval tasks.  

\subsection{Datasets}
\subsubsection{MIMIC-III database}
\label{sec:mimic-iii} 

The MIMIC-III (Medical Information Mart for Intensive Care, \url{https://mimic.physionet.org/}, \cite{MIMIC-III}) is an openly available database comprising unidentified electronic health records of about 40000 patients. The dataset spans more than a decade, with detailed information about individual patient care. Data includes vital signs, medications, laboratory measurements, observations and notes charted by care providers, fluid balance, procedure codes, diagnostic codes, imaging reports, hospital length of stay, survival data, etc. 

In this project we focus mainly on medical notes in raw text and ICD diagnostic codes of patients. We do not include demographic features, medical test results or measurements, because there will not be such kind of information but only raw text in general retrieval tasks. 

In the dataset, each patient's stay is indexed by a \texttt{subject\_id} and a \texttt{hadm\_id}: \texttt{subject\_id} is unique to a patient and \texttt{hadm\_id} is unique to a patient's hospital stay. There are in total $58328$ unique (\texttt{subject\_id}, \texttt{hadm\_id}) pairs. We use the medical notes and diagnostic ICD codes as the dataset in our work.  

\begin{figure}[!htb]
  \includegraphics[width=4in]{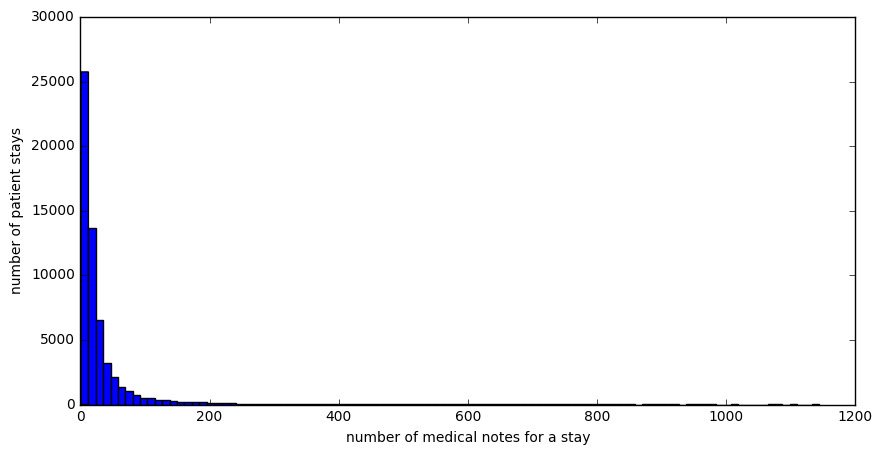}
  \centering
  \caption{Distribution of number of medical notes for each patient's hospital stay.}
  \label{fig:hist-notes-per-sidhid}
\end{figure}

Figure \ref{fig:hist-notes-per-sidhid} shows the distribution of the number of medical notes of each patient's hospital stay. As we can see, most patient's stays have less than 200 notes. 

\begin{figure}[!htb]
  \includegraphics[width=4in]{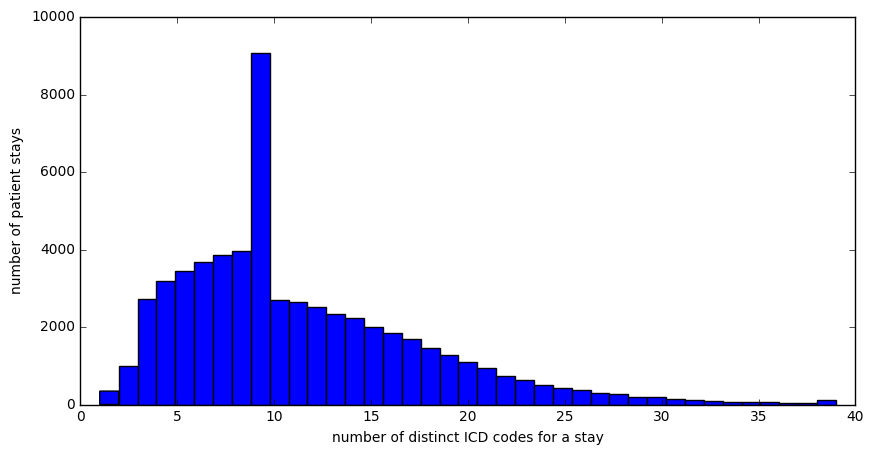}
  \centering
  \caption{Distribution of number of unique diagnostic ICD codes as per patient's hospital stay.}
  \label{fig:hist-icd-per-sidhid}
\end{figure}

As there are too many diagnostic ICD codes, and the distribution is very skewed with a long tail, we are going to take the 50 most frequent ones, which can already cover $93.38$ percent of all patient's stays. 

\subsubsection{TREC Clinical Decision Support Track (2016) dataset} 
\label{sec:trec-cds} 

The clinical decision support track (CDS) of Text REtrieval Conference (TREC) investigates techniques for linking medical records to information relevant for patient care. The 2016 TREC Clinical Decision Support Track (\url{http://www.trec-cds.org/2016.html}) focuses on the retrieval of biomedical articles relevant for answering generic clinical questions about medical records. In this task, the query documents are just admission notes from the MIMIC-III database. The query includes the patient's chief complaint, relevant medical history and other results of tests, all represented in raw text. The task is to retrieve full-text of biomedical articles that might answer questions regarding the query document. 

The candidate document collection is the open access subset of the PubMed Central (PMC, \url{https://www.ncbi.nlm.nih.gov/pmc/}). PMC is an online digital database of freely available full-text biomedical literature. This document collection contains 1.25 million articles in NXML format. 

\subsubsection{RCV1 dataset}
\label{sec:rcv1}

Reuters Corpus Volume I (RCV1) \cite{lewis2004rcv1} is a text classification dataset, which contains over $800, 000$ categorized news stories. There are three types of category codes: \texttt{topics}, \texttt{industries}, and \texttt{regions}. 

The \texttt{topics} codes capture the subject of a news story, \texttt{industries} codes indicate the type of business discussed in the story, and \texttt{regions} codes include both geographical and economical/political groupings. Both \texttt{industries}, and \texttt{regions} codes have a hierarchical structure: there are large and coarse groups of topics/industries and under each large topic/industry there are several finer-grained codes. 

In our experiment, we make use of the \texttt{topics} codes as target labels for embedding model training. As this label is most related to the semantics of an article. We use the fine-grained topics labels, there are in total $126$ of them. 

Another thing to notice is that each new story does not necessarily have just one topic label, which suggests we should not use softmax as final layer activation. This scenario is very similar to the MIMIC-III case, where each patient's stay does not necessarily have just one diagnostic ICD code. 

\subsubsection{Tipster dataset}
\label{sec:tipster}
Tipster \cite{callan1993evaluation} is a large and heterogeneous document collection for information retrieval research, it contains about $750, 000$ newspaper articles, magazine articles, Federal Register announcements and technical abstracts. 

Instead of short queries with just a few words, the query sets in Tipster are described by \emph{topics}. Inside each topic, different aspects of the query is expressed. Each \texttt{topic} consists of several fields: \texttt{domain}, \texttt{title}, \texttt{description}, \texttt{summary}, \texttt{narrative}, \texttt{concepts}, \texttt{factors} and \texttt{definitions}. 

The \texttt{domain} and \texttt{title} provide just a few keywords of the information needed, \texttt{concepts} provides a longer list of relevant keywords, while the fields \texttt{description}, \texttt{summary}, \texttt{narrative} are natural language descriptions of the conditions that make a document relevant to the corresponding topic, which makes the query documents quite long in total. If we consider the fields of \texttt{description}, \texttt{summary}, \texttt{narrative}, the query can be seen as a piece of text that contains several paragraphs. 

Here is an example of query narrative in Tipster: 

\textit{A relevant document will mention the signing of a contract or preliminary
agreement , or the making of a tentative reservation, to launch a commercial
satellite.
}

While the Tipster queries are lengthy text paragraphs, it differs from the TREC-CDS dataset. In Tipster, the query texts are descriptions of the needed information, describing how a \textit{relevant document} should be like. Whereas in TREC-CDS, the query documents are real medical notes from MIMIC-III dataset, in TREC-CDS the query documents do not describe how a relevant article should look like or should contain which keywords, we judge a candidate article as relevant when both the query document and article talks about the same medical concepts. 

\subsection{Evaluation metrics}
\subsubsection{Evaluation metrics for ICD code prediction}
\label{ssec:metrics-icd}

As Figure \ref{fig:hist-icd-per-sidhid} shows, each patient has typically more than one diagnostics ICD codes, we are in a multi-label classification scenario. The same is true when training embedding model on RCV1 dataset. To evaluate the quality of our model in such a scenario, we use the metrics as follows \cite{journals/cleiej/ChermanMM11}: 

Suppose there are $N$ instances to predict and $K$ possible labels for each instance, denote the model predictions as a $N \times K$ matrix $Y$ (row $Y_i \in \{0,1\}^K$ is the prediction for instance $i$) and denote truth labels as $N \times K$ matrix $Z$ (each row $Z_i \in \{0,1\}^K$). The \textbf{precision}, \textbf{recall}, \textbf{F1 score} and \textbf{accuracy} is defined in Equation \ref{eq:metric-multilabel-clf}.

\begin{equation}
\label{eq:metric-multilabel-clf}
\begin{split}
Precision(Y,Z) &= \frac{1}{N}\overset{N}{\underset{i=1}{\sum}}\frac{\left |Y_i\cap Z_i\right |}{\left | Z_i \right |}   \\
Recall(Y,Z) &= \frac{1}{N}\overset{N}{\underset{i=1}{\sum}}\frac{\left |Y_i\cap Z_i\right |}{\left | Y_i \right |}  \\
F(Y,Z) &= \frac{1}{N}\overset{N}{\underset{i=1}{\sum}}\frac{2\left |Y_i\cap Z_i\right |}{\left | Z_i \right |+\left | Y_i \right |} \\
Accuracy(Y,Z) &= \frac{1}{N}\overset{N}{\underset{i=1}{\sum}}\frac{\left |Y_i \cap Z_i\right |}{\left |Y_i \cup  Z_i\right |}  \\
\end{split}
\end{equation}

\subsubsection{Evaluation metrics for retrieval}
For the retrieval task, we will generate ranked list of candidate documents for each query. We use the standard TREC evaluation metrics for ad-hoc retrieval tasks. 

More specifically, 3 metrics are considered in our experiment: Mean Average Precision (\textbf{MAP}), Precision at $20$ (\textbf{P@20}), and Normalized Discounted Cumulative Gain at 20 (\textbf{nDCG@20}). 

Suppose we have $Q$ queries, for each query $q$, $n$ documents are retrieved. Let $\mathbb{I}_{rel(i)}$ be an indicator function equaling $1$ if the $i$th retrieved document is relevant and $0$ otherwise. The precision at $K$ (\textbf{P@K}) is defined in Equation \ref{eq:p-at-k}. 

\begin{equation}
\label{eq:p-at-k}
P@K = \frac{\overset{K}{\underset{i=1}{\sum}}\mathbb{I}_{rel(i)}}{K}
\end{equation}

The average precision for a certain query $q$ (\textbf{$AP_q$}) is defined as a average of \textbf{P@K} for different $K$s. Averaging $AP_q$ over all $Q$ queries gives the mean average precision \textbf{MAP}, as shown in Equations \ref{eq:AP} and \ref{eq:MAP}. 

\begin{equation}
\label{eq:AP}
AP_q = \frac{\overset{n}{\underset{k=1}{\sum}}\mathbb{I}_{rel(k)} \times P@k}{\#relevant\ documents}
\end{equation}

\begin{equation}
\label{eq:MAP}
MAP  = \frac{\overset{Q}{\underset{q=1}{\sum}}AP_q}{Q}
\end{equation}

Let $rel_i$ be the relevance score of $i$th retrieved document ($rel_i \in \{0,1,2\}$ in our test qrels). To encourage relevant documents being ranked higher in the result rankedlist, define Discounted Cumulative Gain at $K$ (\textbf{DCG@K}) as a discounted sum of rewards at each position $i$ from $1$ to $K$, as shown in Equation \ref{eq:dcg-k}. Denote the maximum possible DCG at $K$ as $IDCG@K$, the normalized version of DCG (\textbf{nDCG@K}, Equation \ref{eq:ndcg}) is a measure between $0$ and $1$ thus more comparable. 

\begin{equation}
\label{eq:dcg-k}
DCG@K = \overset{K}{\underset{i=1}{\sum}}\frac{2^{rel_i}-1}{\log(i+1)}
\end{equation}

\begin{equation}
\label{eq:ndcg}
nDCG@K = \frac{DCG@K}{IDCG@K}
\end{equation}

\subsection{Summary}
In this section, we introduce the datasets used in our experiments. The TREC-CDS dataset is our primary goal. We also try our model on more general retrieval tasks such as Tipster. Both TREC-CDS ans Tipster dataset have the particularity that the queries are long, and contain not just keywords but paragraphs of natural language. 

To train a good embedding, we try to use text corpus that are similar to document sets in retrieval tasks, the MIMIC-III is a good choice for TREC-CDS task because the corpus is related to patient care as what articles in PMC dataset are about. And both RCV1 and Tipster document set contain news stories. 

Both MIMIC-III and RCV1 have the particularity that the text have several labels, suggesting us not to use softmax activation in our neural network output layer. 

For classification evaluation metrics, we use the metrics which are suitable for multi-label classification tasks. For information retrieval, we take 3 commonly-used metrics, MAP, P@20 and nDCG@20, the three metrics combined together reflect well the quality of retrieval results. 

\section{Experiments}
\subsection{Experiments with CNN embedding}
We build our model to predict a patient's diagnostic ICD codes. To reduce the skewness of data, we predict the top 50 most frequent ICD codes (which covers about $93$ percent of all patient's stays). The output of our model is a vector of probabilities that the input patient has a corresponding diagnostic ICD code. We then simply choose $0.5$ as threshold, any probability greater than $0.5$ will be predicted as a $1$. 

As described in Section \ref{ssec:metrics-icd}, we use the precision/recall/F1 score for multi-label classifications to evaluate the performance. As the top 50 most frequent ICD codes are still quite skewed, we have modified the sampling weight during the model training to have a bigger sample weight for instances that contains a less frequent ICD code. 

As baseline algorithms, we use the tf-idf representation of patients' notes fed to a SVM algorithm, also we implement the deep patient embedding algorithm as described in Section \ref{ssec:deep-patient}. The deep patient feature vectors are also fed into a SVM. A grid search is used to find the best parameters for SVM. 

To evaluate the quality of the embedding vector itself, we also feed our CNN embedding vectors into a SVM. The experiment results are shown in Table \ref{table:experiment-icd-prediction}. 

\begin{table}[!htb]
\centering
\caption{Results of ICD code prediction from raw medical notes for different algorithms}
\label{table:experiment-icd-prediction}
\begin{tabular}{@{}l|llll@{}}
\toprule
Algorithm & SVM+tfidf & SVM+deep-patient & CNN              & SVM+CNN-embedding \\ \midrule
Precision & 0.1129    & 0.1490           & \textbf{0.4069}  & 0.2162            \\ 
Recall    & 0.5170    & 0.5520           & 0.2335           & \textbf{0.7732}   \\ 
F1 score  & 0.1747    & 0.2228           & 0.2676           & \textbf{0.3104}   \\ 
Accuracy  & 0.1034    & 0.1374           & 0.2001           & \textbf{0.2072}   \\ \bottomrule
\end{tabular}
\end{table}

From the table we can see that the simple tf-idf feature performs the worst among all algorithms, while SVM with our embedding model outperforms all other algorithms. The recall is in general not as good as expected due to the skewness of the ICD codes. 

\subsection{Experiments of information retrieval}
For information retrieval tasks, we test our embedding applied to DRMM on two datasets: one is the 2016 TREC-CDS dataset, the other is the Tipster dataset. To train our embedding, for TREC-CDS we use MIMIC-III dataset as described in previous section, we train the embedding on RCV1 dataset for Tipster. 

The algorithms used are as follows: the traditional BM25 algorithm (\textbf{BM25}) as a baseline, the original DRMM using word vectors (\textbf{DRMM+wd}), our CNN embedding vectors of paragraphs fed to the DRMM model (\textbf{DRMM+para}), and a simple cosine similarity scoring algorithm that calculates the cosine similarity between query and candidate document's embeddings (\textbf{cosine\_sim}). We also deploy an ensemble model taking the sum of scores from the original DRMM, the DRMM+embedding and the cosine similarity as ranking score (\textbf{combined}). 

\subsubsection{Experiments on TREC-CDS dataset} 
Table \ref{table:experiment-ir-trec} shows the result of information retrieval on the TREC-CDS dataset. 

\begin{table}[!htb]
\centering
\caption{Results of retrieval on TREC-CDS2016 dataset for different algorithms}
\label{table:experiment-ir-trec}
\begin{tabular}{@{}l|lllll@{}}
\toprule
Algorithm & BM25    & DRMM+wd & DRMM+para & cosine\_sim     & combined         \\ \midrule
MAP       & 0.1437  & 0.1764  & 0.1812    & 0.1987          &  \textbf{0.2084} \\ 
P@20      & 0.1733  & 0.1783  & 0.1733    & \textbf{0.2267} &  0.2017          \\  
nDCG@20   & 0.1164  & 0.1334  & 0.1198    & \textbf{0.1548} &  0.1510          \\  \bottomrule
\end{tabular}
\end{table}

As we can see, both the original DRMM model and DRMM+embedding outperforms the traditional BM25 model. This is partially due to the fact that in our test, the query is long paragraphs of text, and semantic matching is more important than just exact matching. 

Surprisingly, the simple cosine similarity score performs better than DRMM-based models, which suggests that our embedding model can effectively capture the context semantics of the whole text. 

The ensemble method gives the best MAP score, but the cosine similarity scoring performs the best for metrics P@20 and nDCG@20. 

\subsubsection{Experiments on Tipster dataset}
Table \ref{table:experiment-ir-tipster} shows the information retrieval result on Tipster dataset. 

\begin{table}[!htb]
\centering
\caption{Results of retrieval on Tipster dataset for different algorithms}
\label{table:experiment-ir-tipster}
\begin{tabular}{@{}l|lllll@{}}
\toprule
          & BM25    & DRMM+wd & DRMM+para & cosine\_sim & combined         \\ \midrule
MAP       & 0.1791  & 0.3361  & 0.2732    & 0.2818      & \textbf{0.3573}  \\ 
P@20      & 0.0775  & 0.4237  & 0.3012    & 0.3125      & \textbf{0.4488}  \\ 
nDCG@20   & 0.0807  & 0.4335  & 0.3076    & 0.3043      & \textbf{0.4746}  \\ \bottomrule
\end{tabular}
\end{table}

The DRMM model still performs better than BM25, but on Tipster the original DRMM (employing term-level interactions) gives better results. The cosine similarity score again outperforms the DRMM+embedding model, suggesting that an embedding of the whole document might be better suited than embedding paragraphs. 

The ensemble model gives the best results on all three metrics. Adding embedding of texts boost the performance of term-based DRMM even further.  

\section{Conclusion}
In this paper, we propose an embedding model for pieces of text and show how to plug our new embedding into information retrieval frameworks. Although our embedding is trained primarily for predicting a certain label from text, we use it mainly for information retrieval tasks. We do the extra text classification step because of the lack of sufficient training qrels for our model. 

Empirical results show that the trained model generates a good embedding vector for label prediction. And our experiments show that this embedding vector works well for information retrieval tasks. More specifically, the embedding can provide a very good feature for contextual semantics. When plugging the embedding vector into information retrieval models, the retrieval performance is boosted accordingly.

Experiments also validate our assumption that a good model for text label prediction tasks can generate good embedding vectors for retrieval tasks as well. And our experiments indicate that the embedding gives better as queries become longer, which shows our model might be good for long, contextual queries.

Further exploration directions in the same spirit includes trying out other structures of prediction model, for example, the recurrent neural network. Moreover, given a sufficiently large and general labeled text corpus, we can even train a universal embedding model, and apply the model to whatever piece of text. To make full use of the embedding vector's potentials in information retrieval tasks, we can also try out more IR models with our embedding. 

\section{Acknowledgements}
This research is funded by the Swiss National Science Foundation (SNSF) Ambizione Program under grant agreement no.\ 174025.

\makeatletter
\renewcommand{\@biblabel}[1]{\hfill #1.}
\makeatother

\bibliographystyle{plain}
\bibliography{ref}

\end{document}